\documentclass[12pt,preprint]{aastex}

\usepackage{amsmath}	
\usepackage{amssymb}	
\usepackage{multicol}
\usepackage{float}

\usepackage{color}

\newcommand{\be}{\begin{equation}}
\newcommand{\ee}{\end{equation}}
\newcommand{\bea}{\begin{eqnarray}}
\newcommand{\eea}{\end{eqnarray}}
\newcommand{\bal}{\begin{aligned}}
\newcommand{\eal}{\end{aligned}}

\begin{document}

\title{Accretion in Radiative Equipartition (AiRE) Disks}

\author{Yasaman K. Yazdi and Niayesh Afshordi}
\affil{Perimeter Institute for Theoretical Physics, 31 Caroline St. N., Waterloo ON, N2L 2Y5, Canada}

\affil{Department of Physics and Astronomy, University of Waterloo, Waterloo ON, N2L 3G1, Canada}

\email{yyazdi@pitp.ca}
\email{nafshordi@pitp.ca}

\begin{abstract}
Standard accretion disk theory  \citep{ss1973} predicts that the total pressure in disks at typical (sub-)Eddington accretion rates becomes radiation pressure dominated. However, radiation pressure dominated disks are thermally unstable. Since these disks are observed in approximate steady state over the instability time-scale, our accretion models in the radiation pressure dominated regime (i.e. inner disk) need to be modified. Here, we present a modification to the Shakura \& Sunyaev model, where radiation pressure is in equipartition with gas pressure in the inner region. We call these flows Accretion in Radiative Equipartition (AiRE) Disks. We introduce the basic features of AiRE disks and show how they modify disk properties such as the Toomre parameter and central temperature. We then show that the accretion rate of AiRE disks is limited from above {\rm and} below, by Toomre  and nodal sonic point instabilities, respectively. The former leads to a strict upper limit on the mass of supermassive black holes as a function of cosmic time (and spin), while the latter could explain the transition between hard and soft states of X-ray binaries. 
\end{abstract}


\section{Introduction}\label{intro}

Since the early years after \citet{ss1973} introduced thin disks, they have been known \citep{ss1976, piran1978} to be fraught with thermal instabilities. The instability occurs in the inner regions of the disk, where the pressure is dominated by radiation pressure and the opacity is mostly due to Thomson scattering, resulting in a much stronger temperature dependence in the heating of the disk compared to its cooling. \\

Decades later, the resolution of thermal instabilities still remains one of the major outstanding problems in understanding thin and slim disks. One of the major uncertainties in the Shakura \& Sunyaev (SS) $\alpha$ disk model is the assumption
that the viscous stress is proportional to the total pressure. Early attempts to model thermal (and viscous) stability, such as the works of \citet{sakimoto, stella, merloni}, explored the possibility that the viscous stress might instead be only proportional to the gas pressure. These disks were called $\beta$ disks, where $t_{r\phi}=\beta\, p_{gas}$.  Recent numerical simulations \citep{jiang2013,bran1995,stone1996, mishra2016, 2016MNRAS.459.4397S, 2016MNRAS.462..960S} see the presence of thermal instabilities, where the onset of thermal instability causes the disk to expand or collapse at the time scale of only a few orbits. These local simulations do not see evidence for such $\beta$ disks. \\

Radiative MHD simulations such as those in \citep{2016MNRAS.459.4397S, 2016MNRAS.462..960S} find stable radiatively efficient and strongly radiation pressure dominated disks, in the presence of strong magnetic fields. This has led to the claim that strong magnetic fields could stabilize disks against thermal instabilities \citep{2016MNRAS.459.4397S, 2016MNRAS.462..960S, begelman07, oda09}. If there is not enough magnetic flux, however, the instability once again sets in and a different means for stabilization must be sought. An iron opacity bump has also been suggested by \citet{jiang2016} as a means to postpone (but not avoid) thermal instabilities. In this paper, we present Accretion in Radiative Equipartition (AiRE) disks as an alternative solution to the thermal instability problem in thin and slim disks.

\section{Slim Disk Equations}
In slim (and thin) accretion disks, the main approximations made are that the disk is axisymmetric, stationary, and $h<r$. The background metric is assumed to be Kerr. \\

Let us define the following expressions involving the black hole spin:
\begin{align}
\Delta&=M^4 (r_*^2-2r_*+a_*^2),\\
A&=M^4 (r_*^4+r_*^2a_*^2+2r_*a_*^2),\nonumber\\
\mathcal{C}&=1-3r_*^{-1}+2a_*r_*^{-3/2},\nonumber\\
\mathcal{D}&=1-2r_*^{-1}+2a_*^2r_*^{-2},\nonumber\\
\mathcal{H}&=1-4a_*r_*^{-3/2}+3a_*^2r_*^{-2}\nonumber,
\end{align}
where $a_*=a/M$ and $r_*=r/M$. When quantities have been vertically integrated, a polytropic equation of state has been assumed: $p=\mathcal{K}\ \rho^{1+1/N}$ where $N=3$, and $\mathcal{K}=const$. In the notation that follows, $\Sigma, T_c,$ and $P$ are vertically integrated quantities (whereas $p$, for example, is not). Unless otherwise noted, we use geometrized units, $G=c=1$. $\dot{M}$ is the mass accretion rate and we use the convention that $\dot{M}_{\rm Edd}=16 L_{\rm Edd}/c^2$ \footnote{A common alternative definition of the Eddington mass accretion rate is $\dot{M}_{\rm Edd}=L_{\rm Edd}/c^2$, but here we follow the convention of \citet{sadowski11}.}.\\

Following the notation of \citet{sadowski11, abram11}, the relativistic equations describing slim disks are:

\begin{enumerate}

\item Conservation of (rest) Mass:

\be
\dot{M}=-2\pi\Sigma\sqrt{\Delta}\frac{v}{\sqrt{1-v^2}},
\label{mc}
\ee
where $v$ is the radial velocity and $\Sigma$ is the surface density.\\

\item Conservation of radial momentum:
\be
\frac{v}{1-v^2}\frac{dv}{dr}=\frac{\mathcal{A}}{r}-\frac{1}{\Sigma}\frac{dP}{dr},
\label{rmc}
\ee
where
\be
\mathcal{A}=-\frac{MA}{r^3\Delta \Omega_k^+ \Omega_k^-}\frac{(\Omega-\Omega_k^+)(\Omega-\Omega_k^-)}{1-\tilde{\Omega}^2\tilde{r}^2},
\ee
$\tilde{r}=A/(r^2\sqrt{\Delta})$, and $\Omega$ is the angular velocity with respect to the stationary observer, while $\tilde{\Omega}=\Omega-\frac{2Mar}{A}$ is the angular velocity with respect to the inertial observer. $\Omega_k^\pm=\pm \sqrt{M}/(r^{3/2}\pm a\sqrt{M})$ are the angular velocities of co-rotating and counter-rotating Keplerian (or circular geodesic) orbits. \\

\item Conservation of angular momentum:
\be
\frac{\dot{M}}{2\pi} (\mathcal{L}-\mathcal{L}_{in})=\frac{\sqrt{A \Delta}\gamma}{r}\alpha P,
\label{cam}
\ee
where $\mathcal{L}$ is the angular momentum per unit mass, $\mathcal{L}_{in}$ is an integration constant (approaching the angular momentum at the innermost stable circular orbit (ISCO), in the thin disk limit), and 
\be
\gamma=\sqrt{\frac{1}{1-v^2}+\frac{\mathcal{L}^2r^2}{A}},
\label{lorentz}
\ee
is the Lorentz factor. We adopt the $\alpha$ viscosity prescription, where the $t_{r\phi}$ component (in the co-moving frame of the fluid) of the viscous stress tensor can be expressed as $t_{r\phi}=-\alpha\,p$ \citep{ss1973}. The relation between $\Omega$ and $\mathcal{L}$ is
\be
\Omega=\frac{2 a  M r}{ A}+\frac{ r^3 \mathcal{L} \sqrt{\Delta}}{A^{3/2} \gamma}.
\ee
\item Vertical equilibrium:
\be
h^2\Omega_\perp^2=9\frac{P}{\Sigma},
\label{ve}
\ee
 where $ \Omega_\perp=\sqrt{\frac{M}{r^3}\frac{\mathcal{H}}{\mathcal{C}}}$ is the vertical epicyclic frequency, and $h$ is the half disk thickness.

\item Conservation of energy:
\be
\mathcal{Q}^{\rm adv}=\mathcal{Q}^{\rm vis}-\mathcal{Q}^{\rm rad}.
\label{en_cons}
\ee

Here, the advective cooling is defined as:
\be
\mathcal{Q}^{\rm adv}=\frac{1}{r}\frac{d}{dr}\left[r u^r(E+P)\right]-u^r\frac{dP}{dr}-\int^{h}_{-h} u^z\frac{\partial p}{\partial z}dz,
\ee
where $E$ is the vertically integrated energy density $E=\int^h_{-h}(\frac{3}{2}p_{\rm gas}+3 p_{\rm rad})\ dz$. Performing the $z$ integral in ${Q}^{\rm adv}$, we get

\be
\begin{split}
\mathcal{Q}^{\rm adv}= \frac{\dot{M}}{2\pi r^2}&  \left(\frac{}{}\right.\eta_3\frac{P}{\Sigma}\frac{d\ln P}{d\ln r}-(1+\eta_3)\frac{P}{\Sigma}\frac{d\ln \Sigma}{d\ln r}+\\
& \eta_3 \frac{P}{\Sigma}\frac{d\ln \eta_3}{d\ln r}+\Omega_\perp^2\eta_4 \frac{d\ln \eta_4}{d\ln r} \left. \frac{}{}\right),
\end{split}
\ee
where
\begin{align}
\eta_1&=\frac{128}{315}\,h,\\
\eta_2&=\frac{8}{9},\\
\eta_3&=\frac{1}{P} \left(\frac{1}{5/3-1}\frac{k}{\mu m_p}\frac{8}{9}\Sigma\, T_c+\frac{256}{315}a_{\rm rad}\,T_c^4\,h\right),\\
\eta_4&=\frac{1}{18}\, h^2,
\end{align}
and where k is the Boltzmann constant, $m_p$ is the mass of the proton and $\mu=0.62$ is the solar abundance.

The viscous heating $\mathcal{Q}^{\rm vis}$ is fixed by the $\alpha$-prescription:
\be
\mathcal{Q}^{\rm vis}=-\alpha P \frac{A\gamma^2}{r^3}\frac{d\Omega}{dr},
\label{vis}
\ee
while the radiative cooling $\mathcal{Q}^{\rm rad}$ is: 
\be
\mathcal{Q}^{\rm rad} = \frac{64\sigma T_c^4}{3 \Sigma\kappa},
\label{rad}
\ee
where $T_c$ is the temperature at the equatorial plane, and the opacity coefficient $\kappa$ is given by Kramer's formula \citep{sadowski11}:
 \be
\kappa=\kappa_{es}+\kappa_{ff}=0.34+6.4\times 10^{22}\rho T^{-7/2} 
\label{kap}
 \ee
in cgs units, and using solar abundance. $\rho\sim\frac{\Sigma}{2 h}$ and $T\sim T_c$.
The factor $\frac{64}{3}$ in front of the radiation term in (\ref{rad}) is somewhat arbitrary, as it depends on the details of the assumed vertical structure and radiative transfer, and thus its exact value should be taken with a grain of salt\footnote{We use the coefficient appearing in \citet{sadowski11}, although in other references, such as \cite{sadowski09} it differs by a factor of 2.}.\\

\item The vertically integrated equation of state:
\be
P=\eta_2 \frac{k}{\mu m_p}\Sigma\, T_c+\frac{2}{3}\,\eta_1\, a_{\rm rad}\, T_c^4,
\label{pressure}
\ee
where the first term is the gas pressure $P_{\rm gas}=\eta_2 \frac{k}{\mu m_p}\Sigma\, T_c$, and the second term is the radiation pressure $P_{\rm rad}=\frac{2}{3}\,\eta_1\, a_{\rm rad}\, T_c^4$.
\end{enumerate}
%
 
 In the thin disk (SS) limit of the above equations, Keplerian orbits are assumed ($\Omega \rightarrow \Omega_k$), while  the disk becomes radiatively efficient ($\mathcal{Q}^{\rm adv} \rightarrow 0$). The system then simplifies to an algebraic one where closed form solutions may be found in three regimes: 1) An outer region where $P_{\rm gas}$ and $\kappa_{ff}$ dominate, 2) a middle region where $P_{\rm gas}$ and $\kappa_{es}$ dominate, and 3) an inner region where $P_{\rm rad}$ and $\kappa_{es}$ dominate.

\section{Thermal Instability}
The condition for thermal instability can be written as \citep{pringle}:

\be
\frac{\partial}{\partial h} (\mathcal{Q}^{\rm vis}-\mathcal{Q}^{\rm rad})\Bigr|_\Sigma >0.
\ee

\begin{figure}[h]
\centering
\includegraphics[width=1.1\textwidth]{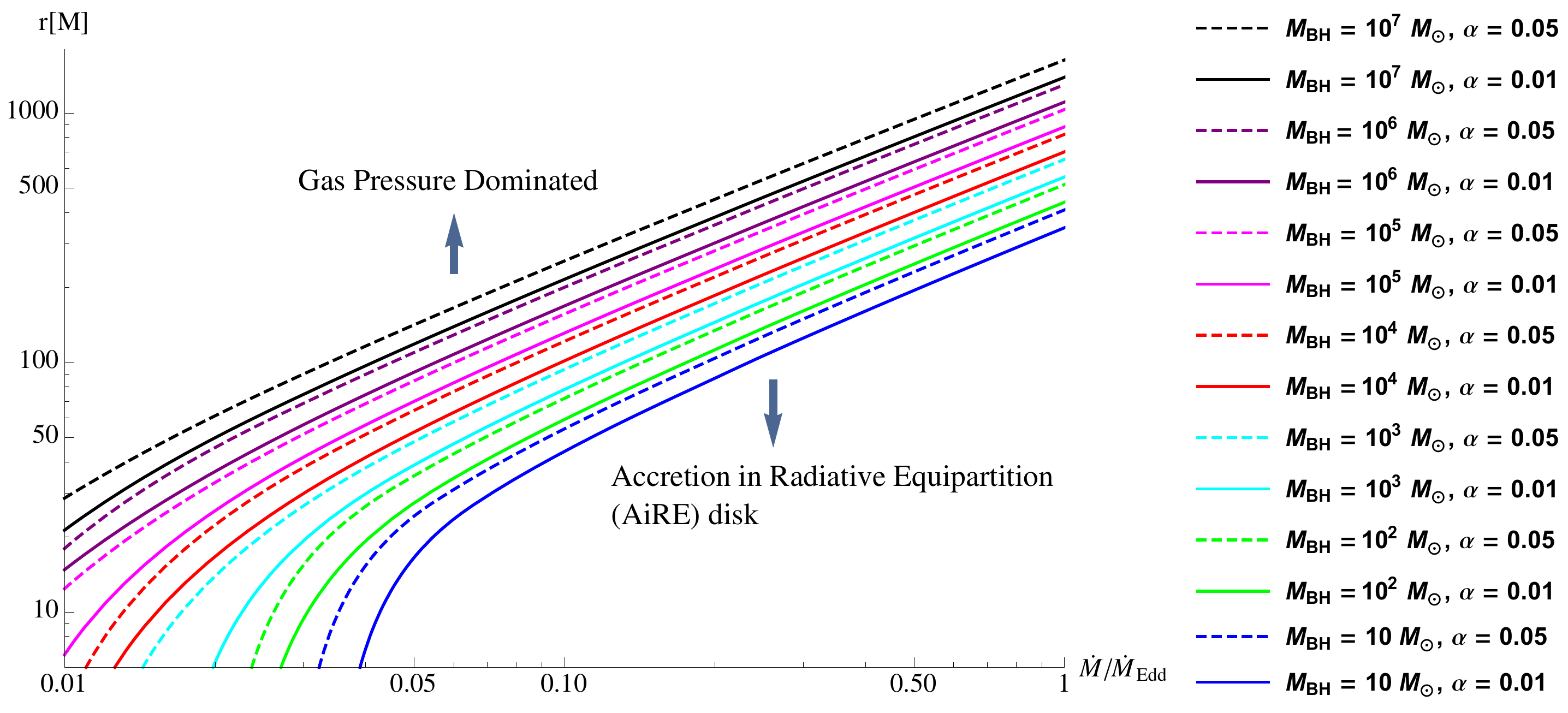}
\caption{Radius at which $P_{rad}=P_{gas}$.}
\label{radii}
\end{figure}

For stellar mass black holes, the slim disks in the previous section are thermally unstable above $\dot{M}/\dot{M}_{Edd}\sim 0.06$, in their radiation pressure dominated regimes. For supermassive black holes, this limit is lower (see Figure \ref{radii}).
The thermal instability sets in because in the innermost regions of the disk we have $p \sim p_{\rm rad}$ and $\kappa \sim$ const., leading to
\be
\mathcal{Q}^{\rm vis}=-\frac{3}{2} \Omega ~t_{r\phi} h \propto p~h \propto h^2,
\ee
where we used vertical equilibrium $p\propto\Omega^2\, \Sigma\, h$ and assumed that $\Sigma$ and $\alpha$ are constant, while
\be
\mathcal{Q}^{\rm rad}=16 \frac{c~ p_{\rm rad}}{\kappa \Sigma}\propto h.
\ee

Since $h\propto p \propto T_c^4$, a temperature increase would change the rate of heating much faster than that of cooling, making the disk thermally unstable.\\

There is also a relation between the shape of the $T_c(\Sigma)$ curve and the stability of the disk. The equilibrium states of the disk can be described by the $T_c(\Sigma)$ solutions, at a fixed radius, to $\mathcal{Q}^{adv}=0$. This $T_c(\Sigma)$ relation is sometimes referred to as an S-curve \citep{abrams, lasota} because of the shape it takes in the models and temperature ranges often considered. Points on this curve with a positive slope correspond to stable equilibria, while points on this curve with a negative slope correspond to unstable equilibria \citep{lasota}. In the former case small perturbations to the temperature bring the state back to  equilibrium, while in the latter case these perturbations lead to runaway heating or cooling.\\

\section{Accretion in Radiative Equipartition}

Figure \ref{radii} shows the radii below which radiation pressure $P_{\rm rad}$ starts to dominate over gas pressure $P_{\rm gas}$ for black holes of different masses and $\alpha$'s. We shall then assume that at this point in the accretion flow, and inwards, the onset of thermal instability creates an inhomogeneous disk structure, where photons can effectively escape faster than they would through diffusion in a smooth disk. In other words, we hypothesize that the effective optical depth (or opacity) of the disk drops due to the instability, or we have more efficient cooling.  This could be e.g. due to rising photon bubbles, low density funnels, or other inhomogeneous structures that could invalidate  \eqref{rad}.\\

We assume that this efficient cooling can cool the disk down to pressure equipartition. However,  the cooling would not  drive the disk to become gas pressure dominated again, as the thermal instability responsible for faster cooling ceases in this regime and the disk heats back up within a viscous time. Therefore, the state of marginal thermal instability, or pressure equipartition, is a stable fixed point (see Sec. \ref{cooling} below for details). Indeed, since disks at high Eddington ratios are observed in a steady state, we conjecture that the cooling is efficient enough to keep the disk at equipartition  i.e. $P_{\rm gas} \approx P_{\rm rad}$\footnote{More generally we can assume $P_{\rm gas} \approx \zeta P_{\rm rad}$. For example, \citet{sadowski11} finds $\zeta = 2/3$ for the onset of thermal instability, but here we consider $\zeta=1$ for simplicity. }. \\

We shall call these flows Accretion in Radiative Equipartition (or AiRE) disks. This equipartition tames the temperature dependence of the disk in the inner region. Within the radii of Figure \ref{radii}, AiRE disks have different properties from slim and thin disks. In this model, the equipartition condition replaces the radiative energy loss condition \eqref{rad}, which is only valid for atmospheres in vertical equilibrium with planar symmetry. More generally, many MHD/fluid instabilities occur when there are different fluid components that dominate inertia and (isotropic or anisotropic) stress. Some examples are Magneto-Rotational Instability (MRI), convective instability, Rayleigh-Taylor instability, Parker instability, and potentially whatever sets the maximum mass of main-sequence stars. In all these, the instability saturates at equipartition, where stress and inertia are (somewhat) equally distributed in dominant components. For example, in gas-dominated disks, MRI saturates when  $P_{\rm mag} = {\cal O}(0.1) \times P_{\rm gas}$ \citep{hawley}. \\

\subsection{Efficient Cooling and Pressure Equipartition}\label{cooling}

Let us now consider how efficient cooling can lead to pressure equipartition. 

\begin{figure}[p]
\centering
\begin{multicols}{2}
\hbox{ \hspace{-.5 cm}\includegraphics[width=.89 \linewidth]{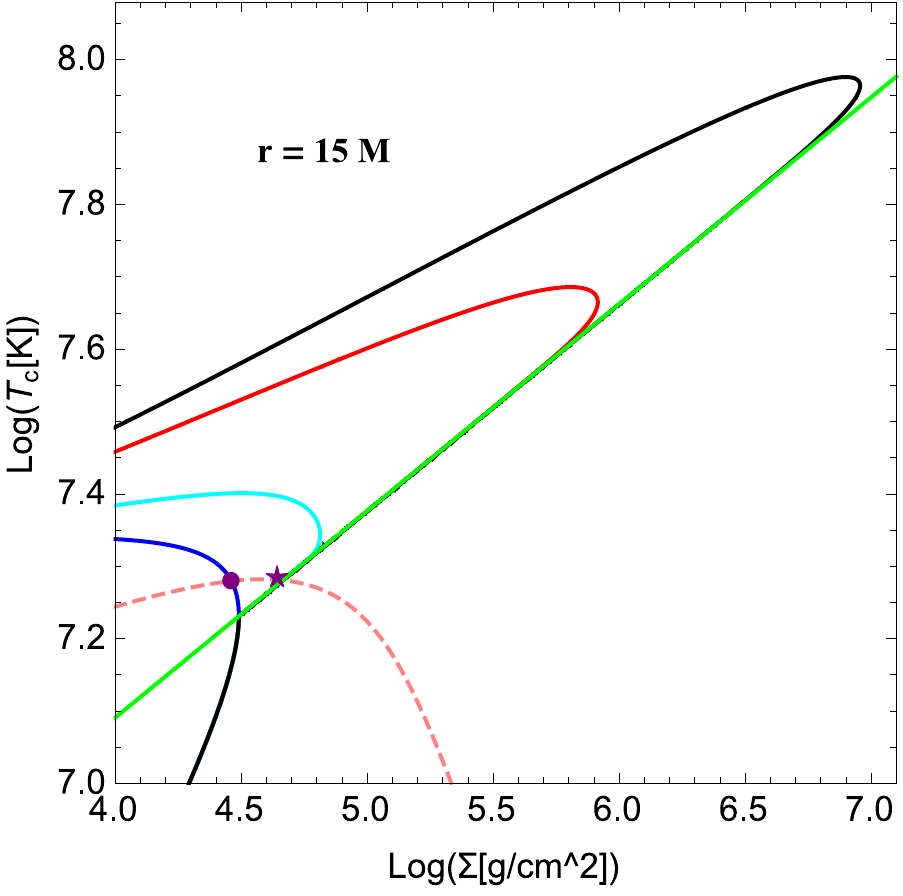}}\par 
\hbox{ \hspace{-1 cm}  \includegraphics[width=1.2 \linewidth]{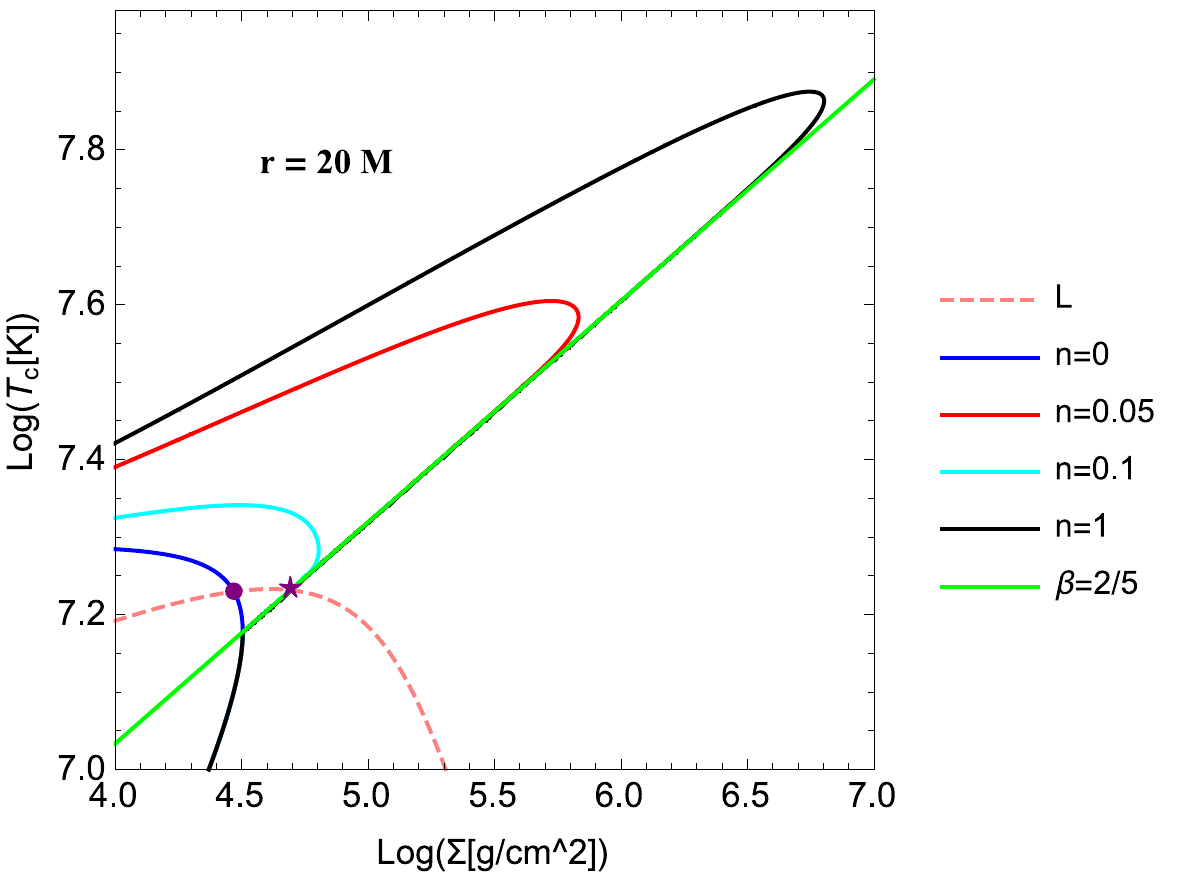}}\par 
    \end{multicols}
\begin{multicols}{2}
   \hbox{\hspace{-.5 cm}\includegraphics[width=.9 \linewidth]{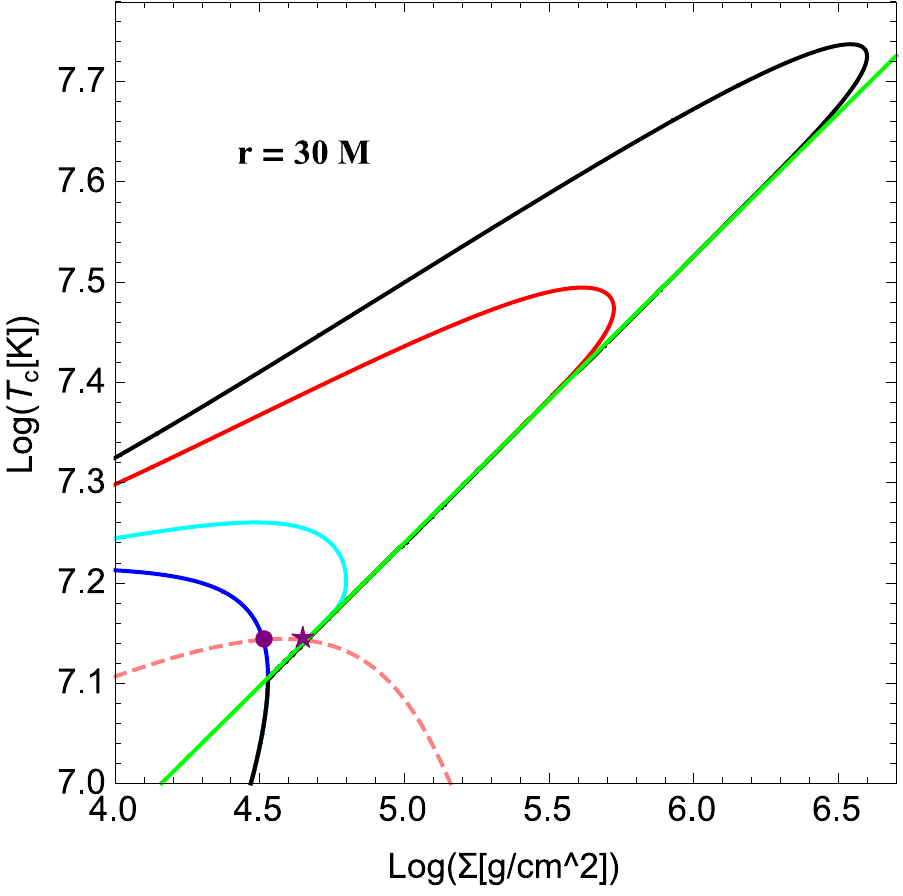}}\par
\hbox{ \hspace{-1 cm}    \includegraphics[width=1.18 \linewidth]{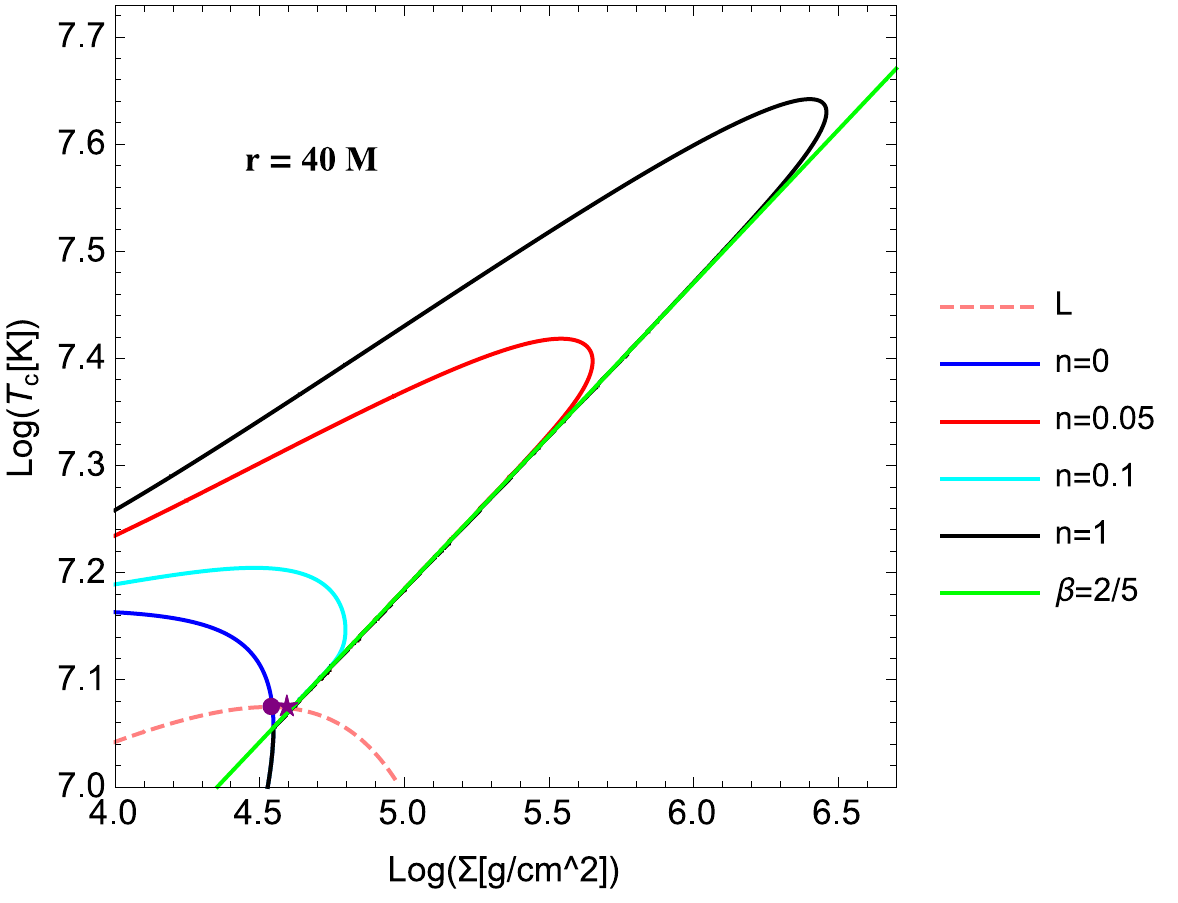}}\par
\end{multicols}
\caption{Thermal equilibria at different fixed radii: $r=15\,M, \, 20\,M,\, 30\,M, \,40\,M$ from top-left to bottom-right. The dashed curve $L$ represents solutions with conserved angular momentum. The intersection of this $L$ curve with the standard model ($n=0$) is marked with a circle and the intersection of this curve with the AiRE disk model ($n>0$) is marked with a star. $\dot{M}/\dot{M}_{Edd}=0.1$ and $\alpha=0.1$.}
\label{thermequib}
\end{figure}

Since the development of inhomogeneities that (could) lead to more efficient cooling is driven by thermal instability, we can write a phenomenological toy model for cooling rate as:
\bea
\tilde{\mathcal{Q}}^{\rm rad}\equiv{\mathcal{Q}^{\rm rad}}\left(1+\text{Re}[\sqrt{2-5\beta}]\frac{t_{vis}}{t_{th}}\right)^{n} \nonumber\\ = \frac{64\sigma T_c^4}{3 \Sigma\kappa}  \left(1+\text{Re}[\sqrt{2-5\beta}]\frac{t_{vis}}{t_{th}}\right)^{n} 
\label{rad_toy}
\eea
where $\beta=\frac{P_{gas}}{P_{gas}+P_{rad}}$, $t_{th}$ and $t_{vis}$ are the thermal and viscous timescales respectively, and $n>0$ is a free parameter. $n=0$ corresponds to the standard disk model, for which $\tilde{\mathcal{Q}}^{\rm rad}\equiv{\mathcal{Q}^{\rm rad}}$ from Eq. (\ref{rad}).  As the rate of thermal instability is $\lambda_{th} \sim  \text{Re}[\sqrt{2-5\beta}]t^{-1}_{th}$, which operates over an accretion/viscous time $\sim t_{vis}$, we expect the cooling to be roughly enhanced by $(\lambda_{th}t_{vis})^n$ with $n= {\cal O}(1)$. Therefore, Eq. (\ref{rad_toy}) gives a reasonable toy model for the instability-driven cooling.  \\

Figure \ref{thermequib} shows the thermal equilibria solutions to $\tilde{\mathcal{Q}}^{adv}\equiv\mathcal{Q}^{vis}-\tilde{\mathcal{Q}}^{rad}=0$ at different fixed radii ($r=15\,M, \, 20\,M,\, 30\,M, \,40\,M$ from top-left to bottom-right) for $n=0, \, 0.05,\, 0.1,\,$ and $1$. The dashed curve $L$ represents solutions with conserved angular momentum. The intersection of this $L$ curve with each equilibrium curve picks out an equilibrium point we are interested in. The intersection with the standard disk model is marked with a circle and the intersection with the AiRE disk model is marked with a star. At larger radii these intersections are closer to one other, but as we move inwards in the disk, the AiRE disk gives us different equilibria compared to the standard model. The slope at the circle is negative, corresponding to an unstable equilibrium, while the slope at the star is positive, corresponding to a stable equilibrium. This confirms our expectation that efficient cooling (\ref{rad_toy}), driven by thermal instability, leads to radiative pressure equipartition $P_{\rm gas} \approx P_{\rm rad}$, which is thermally stable. \\

We think that local shearing box simulations such as \citet{jiang2013} have not yet seen this pressure equipartition realized, either due to insufficient running time or more likely due to the limited box size. Most of the runs in these simulations show either expanding or collapsing vertical height. However, in a big enough box, both should be happening in different places, and this could lead to the inhomogeneities we have just described.\\

\citet{mishra2016}  also see thermal instability in simulations of initially radiation-pressure dominated thin disks.  They argue that a comparison of their evolution with the relevant thin-disk thermal equilibrium curve suggests that their disk may be headed for the thermally stable, gas-pressure-dominated branch, meaning that it is moving towards equipartition. This supports our conjecture. Their simulations, however, had to be terminated before equipartition could occur because they reached a point where they could no longer resolve the disk. 
\subsection{AiRE Disk Equations}

Using $P_{\rm gas} \approx P_{\rm rad}$, which for AiRE disks effectively replaces Eq. (\ref{rad}) in the slim disk model, the equation of mass conservation \eqref{mc} and vertical equilibrium \eqref{ve}, $T_c$ and $v$ can be related:
\be
\dot{M} \approx-2 \pi \sqrt{\Delta} v\frac{\sqrt{18 k/\mu m_p} ~ \xi~ {T_c}^{7/2}}{\Omega_\perp}
\ee
where $\xi\equiv\frac{2}{3}\times \frac{128}{315}\times \frac{a_{\rm rad}}{\eta_2 k/\mu m_p}$.\\

Thus, with the assumption of equipartition of gas pressure with radiation pressure, the equations of the previous section can be combined into one ordinary differential equation, namely the equation of radial momentum conservation \eqref{rmc}. This equation can be rewritten in the form
\be
\frac{d v}{d r}=\frac{\mathcal{N}(r, v)}{\mathcal{D}(r, v)}
\label{ode},
\ee
where

\be
\mathcal{N}(r, v)=\frac{\mathcal{A}}{r}
\ee
and 
\be
\mathcal{D}(r, v)\approx v-1.3 \frac{k}{v\, \mu m_{p}} \left(-\frac{\eta_2 \dot{M} \Omega_\perp  \sqrt{\frac{k}{\mu  m_p}}}{a_{rad}v \sqrt{\Delta } }\right)^{2/7}.
\ee

\begin{figure}[H]
\centering
\includegraphics[width=.9\textwidth]{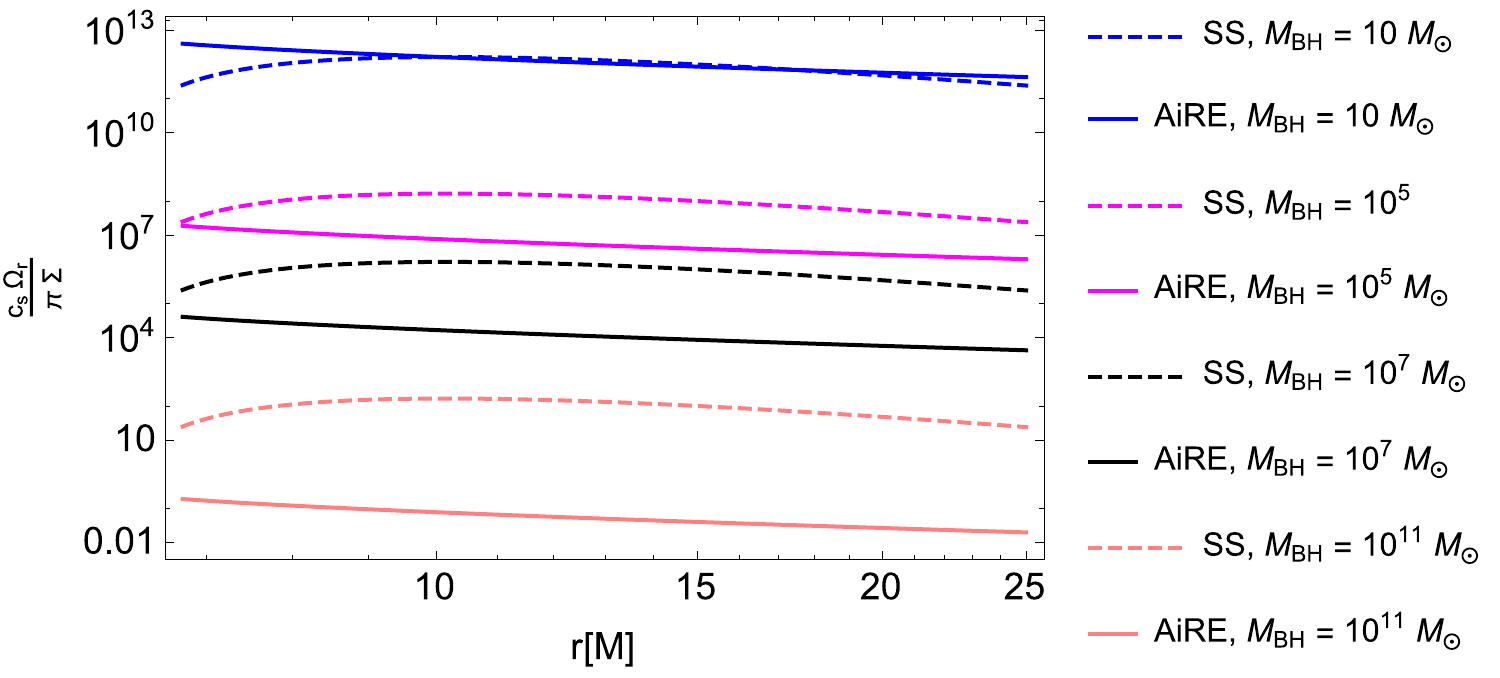}
\caption{Toomre parameter of AiRE disks and SS disks for different black hole masses. $\dot{M}/\dot{M}_{Edd}=0.1$, $a=0$ and $\alpha=0.01$.}
\label{toomres}
\end{figure}

\begin{figure}[H]
\centering
\includegraphics[width=.9\textwidth]{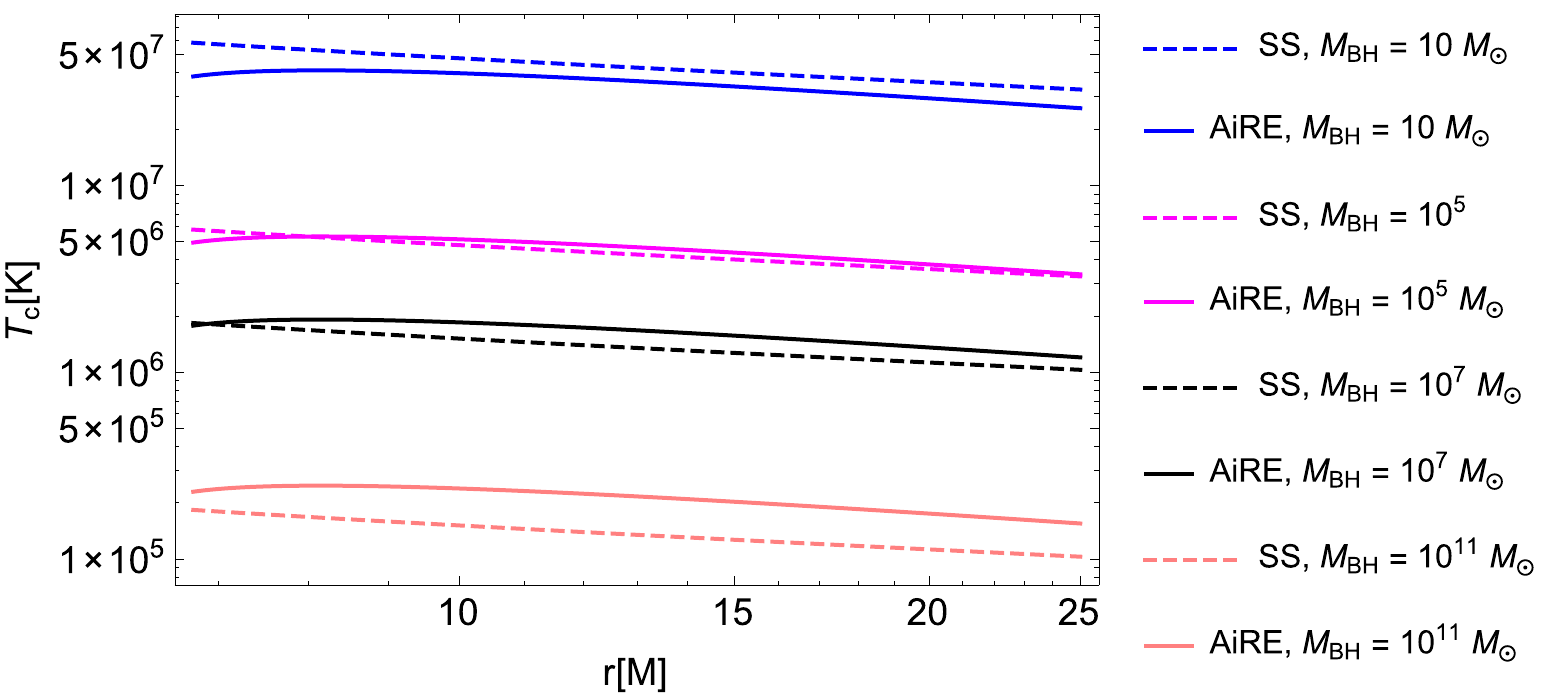}
\caption{Central Temperature of AiRE disks and SS disks for different black hole masses. $\dot{M}/\dot{M}_{Edd}=0.1$, $a=0$ and $\alpha=0.01$.}
\label{tcs}
\end{figure}

Figures \ref{toomres} and \ref{tcs} show the (dimensionless) Toomre parameter and central temperature for AiRE disks in comparison with SS disks, using the Keplerian approximation ($\Omega \rightarrow \Omega_k$) for different non-spinning black hole masses with $\dot{M}/\dot{M}_{Edd}=0.1$, and $\alpha=0.01$. Work is in progress to find the spectrum of the AiRE disks and compare it to the SS spectrum \citep{ykyna} . It would be interesting to see if the AiRE disk spectrum may resolve some of the discrepancies that come from using the SS spectrum, such as in the determination of spins of black holes \citep{abram11}.\\

The thin disk solution in the outer region is used to set the boundary condition far outside the disk. In the innermost  region, the flow must continuously make a transition from subsonic to supersonic flow. This occurs at a radius called the sonic point. The denominator $\mathcal{D}$ in Equation \eqref{ode} vanishes at the sonic point, making the equation singular, unless the numerator $\mathcal{N}$ also vanishes at this point. Thus at the sonic point we must have the inner boundary condition:
\be
\mathcal{N}\Bigr|_{r=r_{sonic}}=\mathcal{D}\Bigr|_{r=r_{sonic}}=0.
\label{nd0}
\ee

The value of the angular momentum constant $\mathcal{L}_{in}$ must be chosen carefully to achieve this. If we call the value of $\mathcal{L}_{in}$ which achieves this $\mathcal{L}_{in}^t$, then for smaller values of this constant $\mathcal{L}_{in}<\mathcal{L}_{in}^t$, $\mathcal{N}$ will vanish but not $\mathcal{D}$ (there will not be a transition from subsonic to supersonic flow), while for greater values of this constant $\mathcal{L}_{in}>\mathcal{L}_{in}^t$, $\mathcal{D}$ will vanish but not $\mathcal{N}$ (there will be a singularity).\\

\begin{figure}[h]
\centering
\includegraphics[width=.75\textwidth]{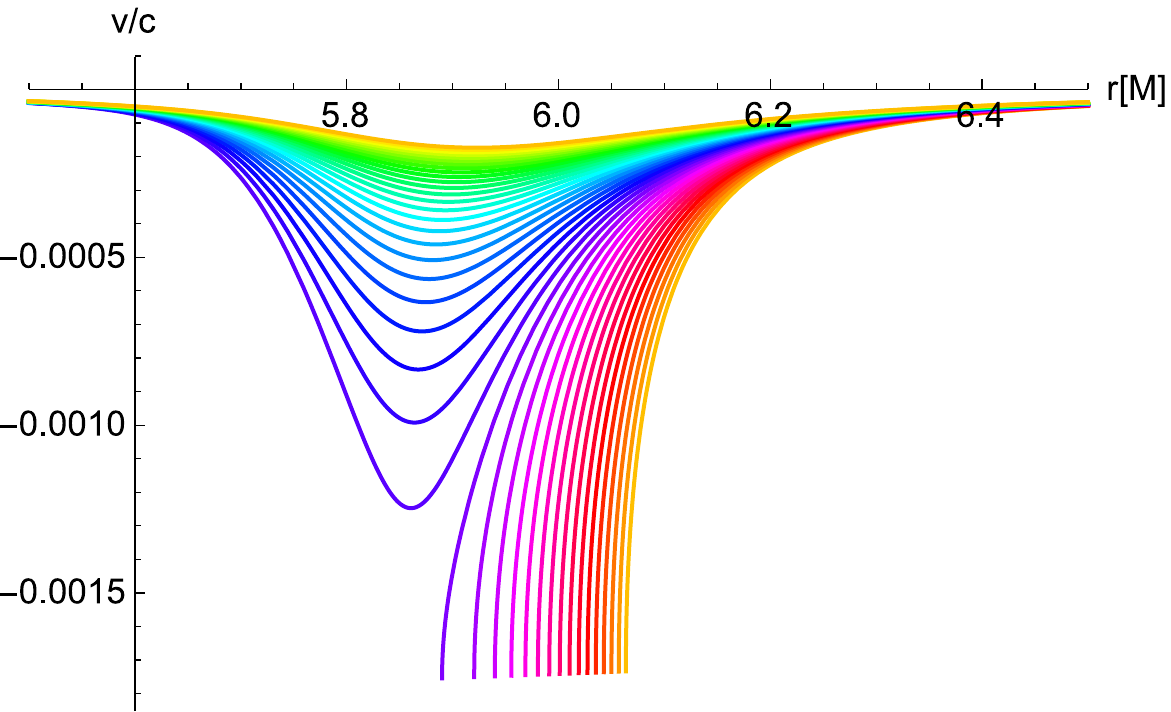}
\caption{Radial velocity profiles for different values of $\mathcal{L}_{in}$. Solutions with a minimum have $\mathcal{L}_{in}<\mathcal{L}_{in}^t$. $M=10\,M_{\odot}$, $\dot{M}/\dot{M}_{Edd}=0.1$, $a=0$ and $\alpha=0.01$.}
\label{vs}
\end{figure}

The radial momentum equation \eqref{ode} may be solved by the shooting method, the relaxation method, or a combination of the two \citep{PressNR}. The topology of $v$ differs for solutions with $\mathcal{L}_{in}>\mathcal{L}_{in}^t$ and $\mathcal{L}_{in}<\mathcal{L}_{in}^t$, and this topology change is key to finding the solution via the shooting method. Figure \ref{vs} shows radial velocity profiles for different values of $\mathcal{L}_{in}$ for $M=10\,M_{\odot}$, $\dot{M}/\dot{M}_{Edd}=0.1$, $a=0$ and $\alpha=0.01$. Solutions with a minimum have $\mathcal{L}_{in}<\mathcal{L}_{in}^t$, whereas the solutions that diverge have $\mathcal{L}_{in}>\mathcal{L}_{in}^t$. The solution with the desired inner boundary condition \eqref{nd0} lies at the transition between these two branches of solutions.\\

 An additional challenge in this problem is that the location of the sonic point is not known in advance. This can be handled by treating the sonic point as a \emph{free boundary}  \citep{PressNR}, when using the relaxation method.

\section{The Nature of the Sonic Point}

\begin{figure}
\centering
\includegraphics[width=.67\textwidth]{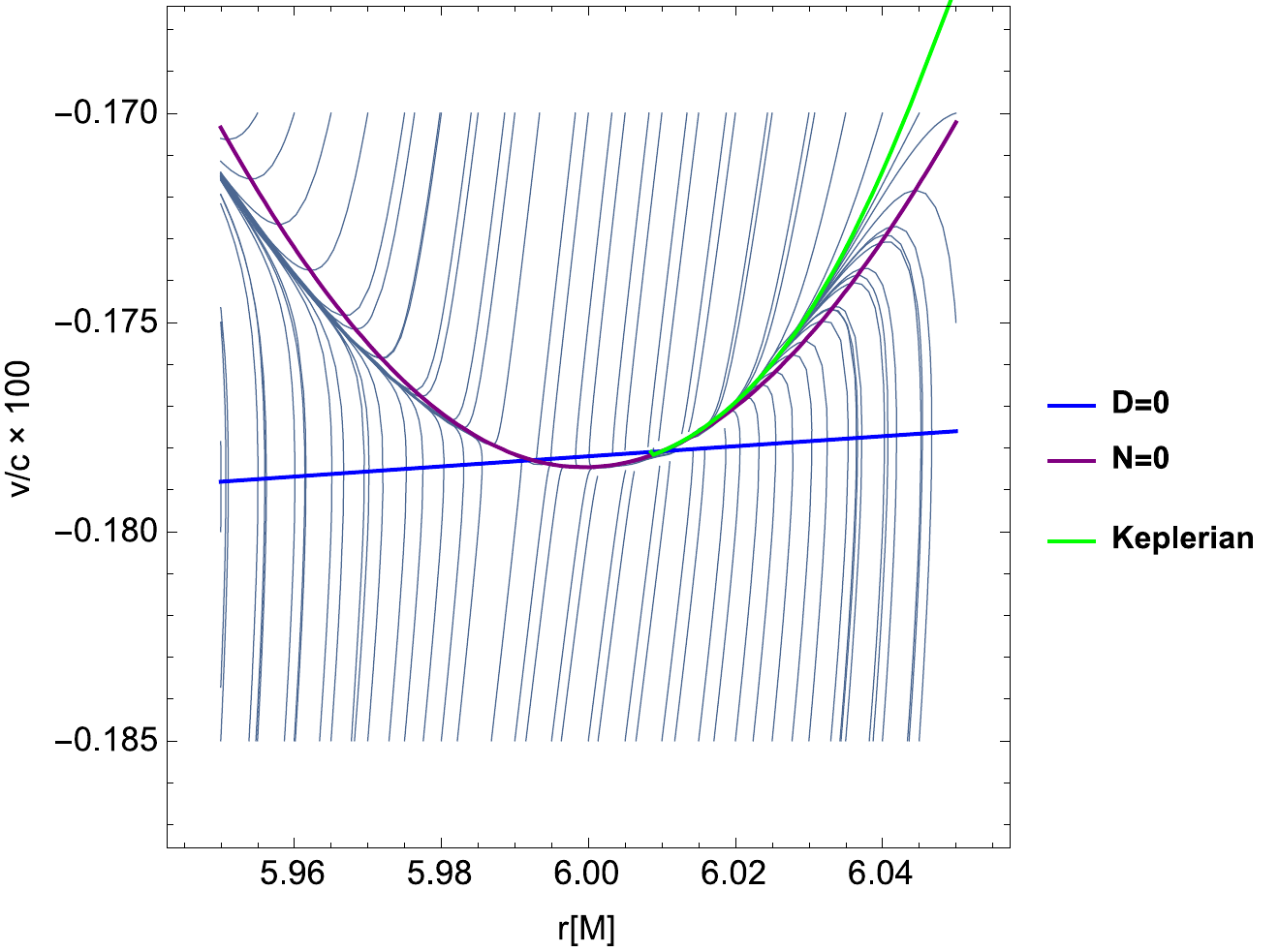}
\caption{Phase portrait for $\dot{M}/\dot{M}_{Edd}=0.12,\, M=10 M_{\odot},\, a=0,\, \alpha=0.2$. There are nodal sonic points at $r\simeq6.01$ and $r\simeq5.99$. The solution with Keplerian outer boundary conditions (in green) goes through the nodal point.}
\label{nod}

\vspace{.8cm}
\hspace{.3cm}\includegraphics[width=.68\textwidth]{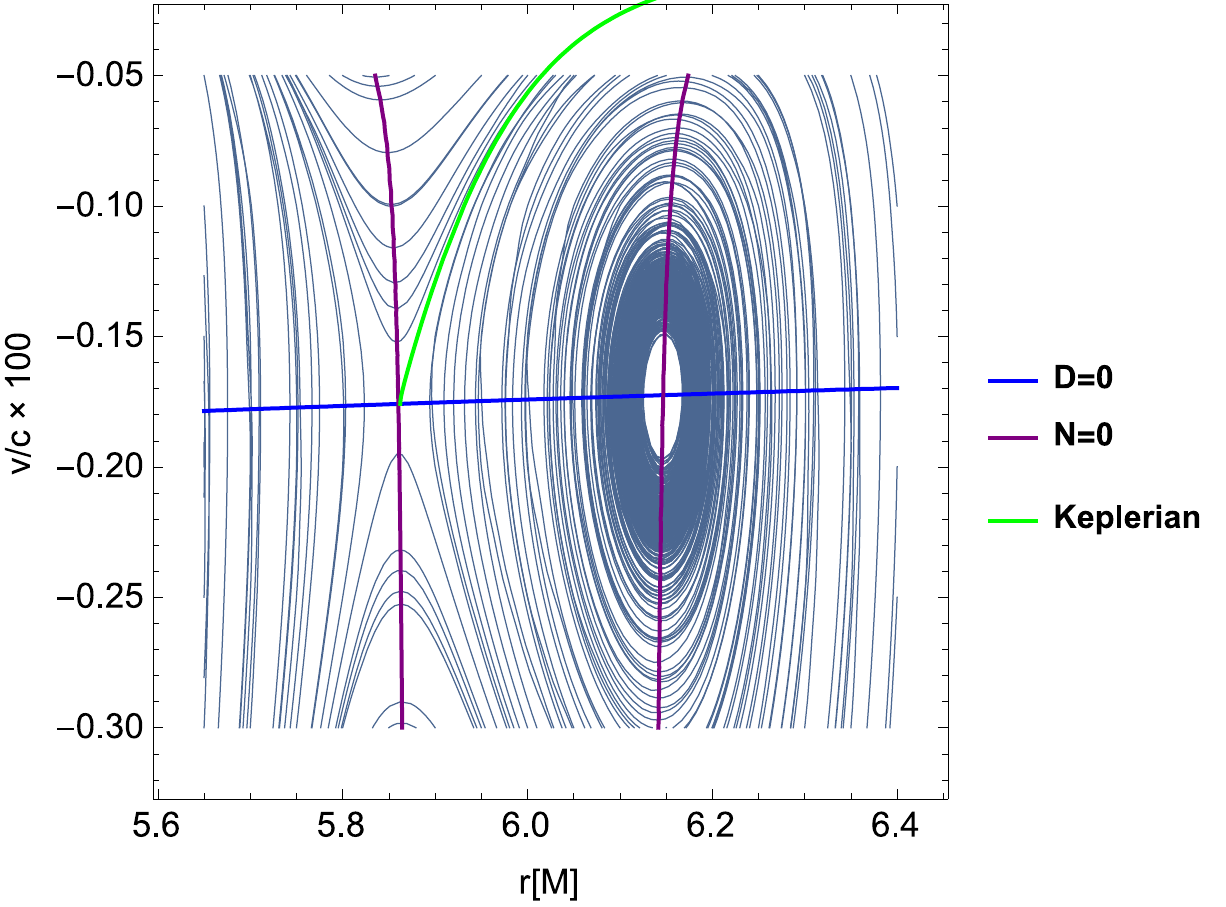}
\caption{Phase portrait for $\dot{M}/\dot{M}_{Edd}=0.1,\, M=10 M_{\odot},\, a=0,\, \alpha=0.01$. There is a Stable saddle point at $r \simeq 5.85$ and an unphysical spiral point at $r \simeq 6.15$. The solution with Keplerian outer boundary conditions (in green) goes through the saddle point.}
\label{sadspir}
\end{figure}

\begin{figure}[H]
\centering
\begin{multicols}{2}
\hbox{ \hspace{-.2 cm}\includegraphics[width=1. \linewidth]{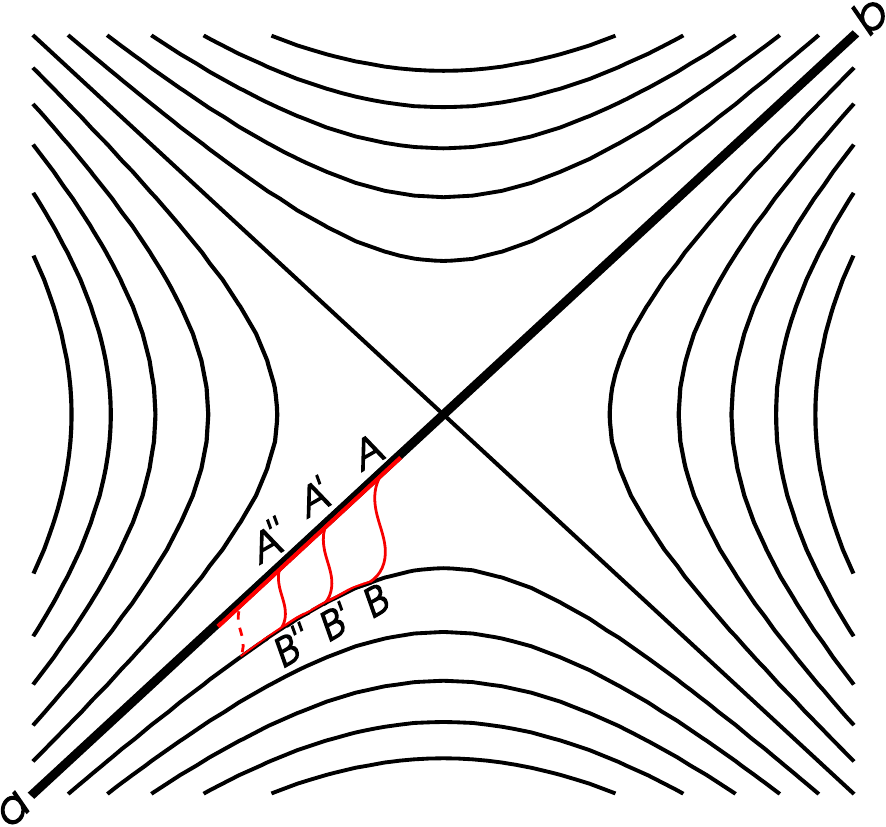}}\par 
\caption{Evolution of perturbations to a steady state solution going through a saddle critical point. The steady state solution is the thick black line $ab$. As perturbations evolve $aBAb\rightarrow aB'A'b\rightarrow aB''A''b ...$, their amplitude shrinks.}
\label{sadpert}
\hbox{ \hspace{0 cm}  \includegraphics[width=.92 \linewidth]{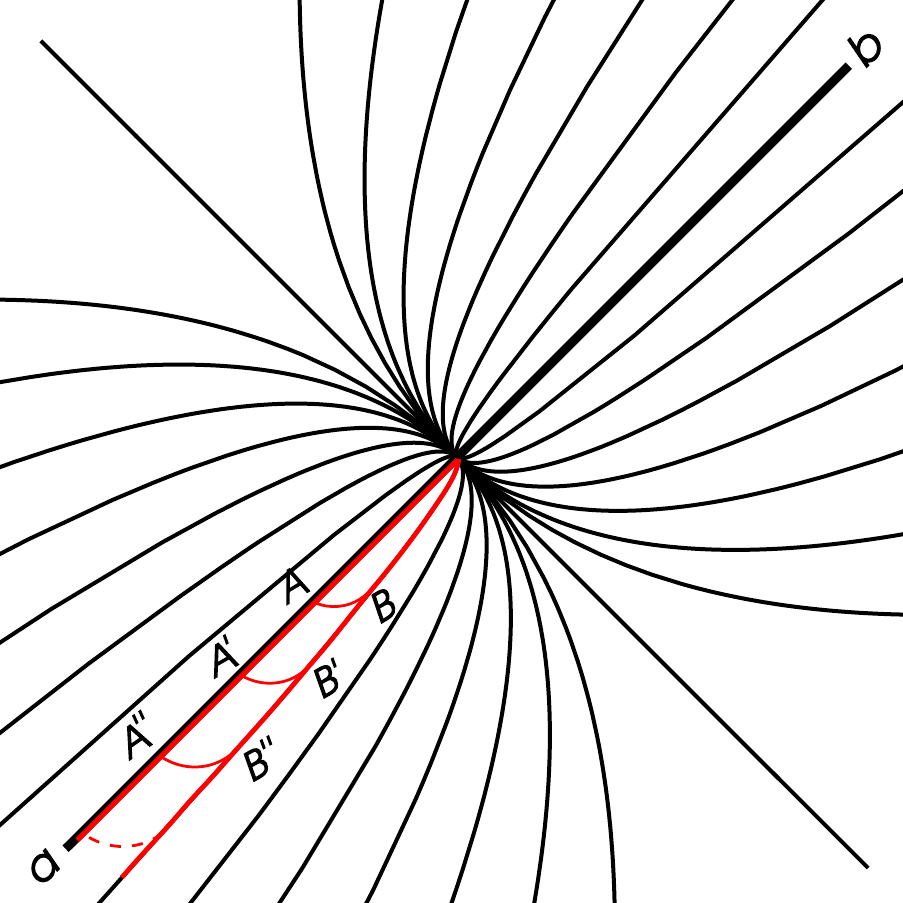}}\par 
\caption{Evolution of perturbations to a steady state solution going through a nodal critical point in the fast direction. The steady state solution is the thick black line $ab$. As perturbations evolve $aBAb\rightarrow aB'A'b\rightarrow aB''A''b ...$, their amplitude grows.}
\label{nodfastpert}
    \end{multicols}
\begin{multicols}{2}
\centering\hbox{\hspace{.2 cm}\includegraphics[width=.94 \linewidth]{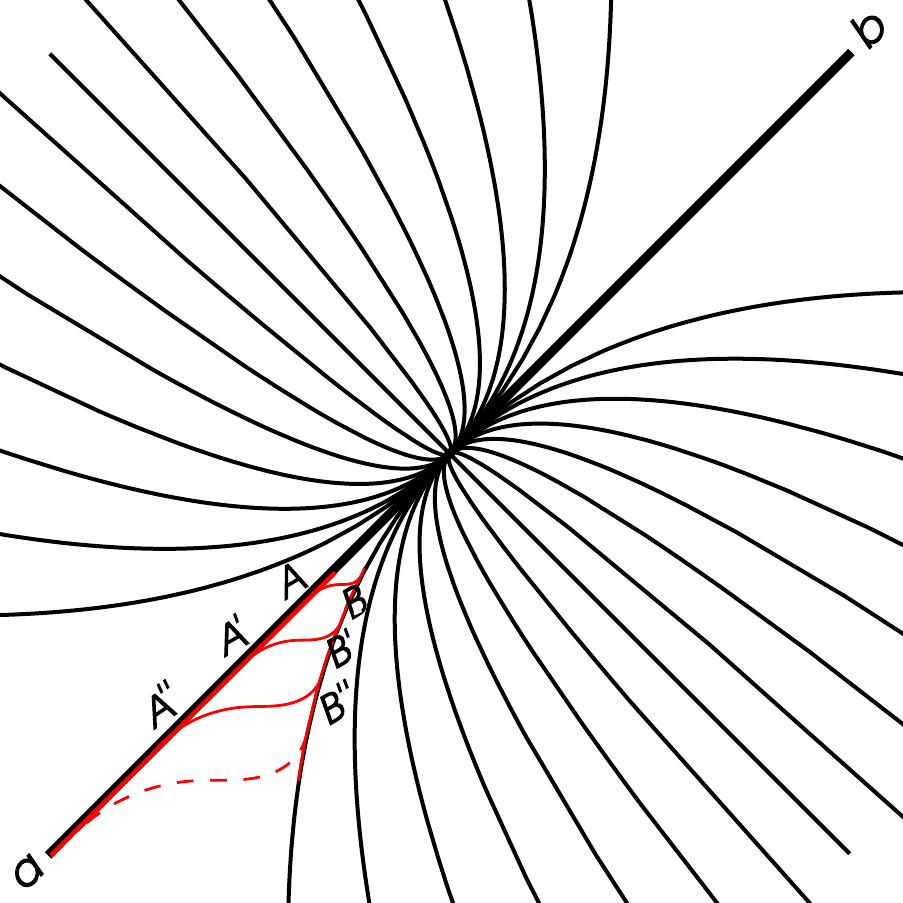}}\par
\caption{Evolution of perturbations to a steady state solution going through a nodal critical point in the slow direction. The steady state solution is the thick black line $ab$. As perturbations evolve $aBAb\rightarrow aB'A'b\rightarrow aB''A''b ...$, their amplitude grows.}
\label{nodslowpert}
    \end{multicols}
\end{figure}

Once we have found the solution with the correct inner boundary condition, we can study the nature of the sonic point by considering the Jacobian matrix $
\mathcal{J}=\begin{pmatrix}
  \frac{\partial\mathcal{D}}{\partial r} &   \frac{\partial\mathcal{D}}{\partial v} \\ 
  
   \frac{\partial\mathcal{N}}{\partial r} &   \frac{\partial\mathcal{N}}{\partial v} 
 \end{pmatrix}
\label{jac}
$. The relative signs of the eigenvalues of this matrix, at the sonic point, characterize the sonic point. If the eigenvalues have the same sign, the sonic point is a nodal point (Figure \ref{nod}), if they have opposite signs, it is a saddle point  (Figure \ref{sadspir}), and if they are complex, the sonic point is a spiral (unphysical) point  (Figure \ref{sadspir}). Given that perturbations can only propagate downstream beyond the sonic point, we conjecture that saddle type sonic points correspond to stable disk configurations, while nodal points correspond to unstable ones. To see this, note that small perturbations inside nodal\footnote{There are two ways of going through a nodal point: in the fast direction (Figure \ref{nodfastpert}) and in the slow direction (Figure \ref{nodslowpert}). Our argument may only hold for passing through the nodal point in the slow direction, as the perturbations are smooth only for this direction.} sonic points grow as they get dragged deeper into the black hole, while small perturbations inside saddle sonic points shrink (see Figures \ref{sadpert}-\ref{nodslowpert}). Other authors have made similar arguments about the instability of nodal sonic points \citep[e.g.,][]{kato1,kato2}. \\

\begin{figure}[h]
\centering
\includegraphics[width=.95\textwidth]{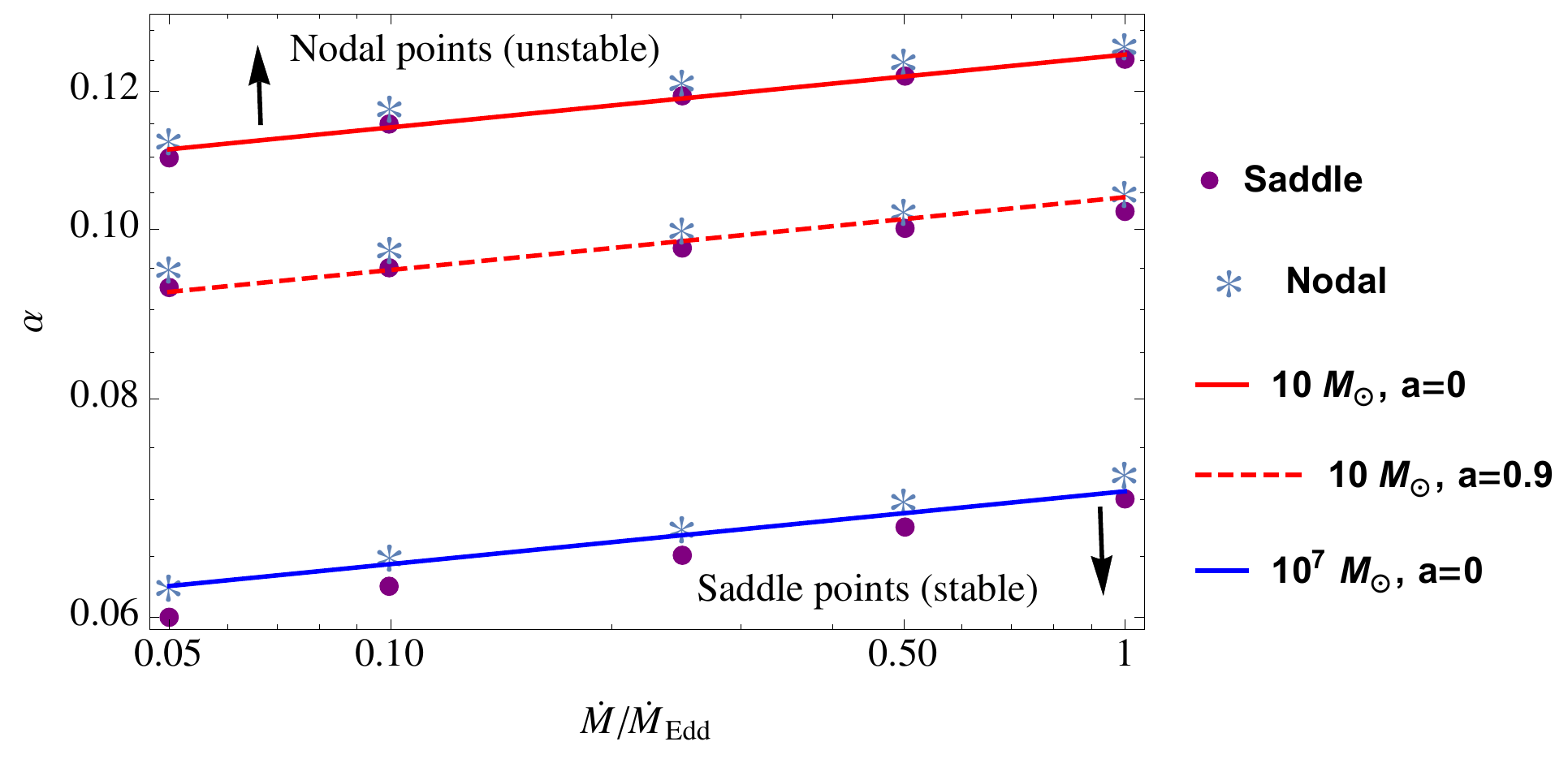}
\caption{Saddle (stable) sonic point to nodal (unstable) sonic point transitions.}
\label{sadnod}
\end{figure}

For fixed values of $\dot{M}/\dot{M}_{Edd},~M$ and $a$, there is a transition from the saddle to the nodal type of sonic point, as we increase $\alpha$. Figure \ref{sadnod} shows some transition lines for different masses and spins of black holes. The transition occurs at 
\be
\begin{split}
\alpha_{\rm saddle} &\lesssim 0.77 \left(-\frac{v_r}{v_\phi}\right)^{1/3} \\
&\simeq 0.097 \left(1-a\right)^{2/15} \left(3+a\right)^{1/3} \left(\frac{\dot{M}/\dot{M}_{Edd}}{M/M_{\odot}}\right)^{1/24}, 
\end{split} \label{alpha_saddle}
\ee
where $v_r=\frac{1}{\gamma} \frac{v}{\sqrt{1-v^2}}$ and $v_\phi=\frac{2 M a}{r\sqrt{\Delta}}+\frac{1}{\gamma}\frac{\mathcal{L} r}{\sqrt{A}}$. Based on our conjecture, we think that values of $\alpha$ greater than $\sim 0.77 \left(-\frac{v_r}{v_\phi}\right)^{1/3}$ correspond to nodal sonic points and are thus unstable, while AiRE disks with smaller $\alpha$'s have healthy saddle sonic points. A similar expression for this transition was found by \citet{afshordi} in pseudo-Newtonian isothermal disks, but with $\frac{h}{r}$ instead of $\frac{v_r}{v_\phi}$. Our expression is more general in that it holds for arbitrary spin in full general relativity.\\

Assuming that the value of $\alpha$ is fixed by the MHD physics in thin accretion disks, Eq. (\ref{alpha_saddle}) suggests a physical origin for the minimum Eddington ratio observed for the soft states of X-ray binaries \citep[e.g., see][]{2013ApJ...779...95K}, which is around 1-3\%. Combining these values with the current measurements of black hole masses and spins \citep[e.g.,][]{2014SSRv..183..295M} gives a range of $\alpha = 0.11-0.13$ for the viscosity parameter.  

We should note that our proposed mechanism for the hard to soft transition would only be consistent with observations if $\alpha$ lies in this narrow range. Current variations of $\alpha$ in simulations are driven by their numerical and/or physical limitations (e.g., thermal instability). However, a better understanding of the physics involved could sharpen these predictions.
In other words, we have a strong prediction for how $\alpha$ should behave in more realistic accretion simulations, but a very weak prediction for the transition Eddington ratios (in the absence of a precise value of $\alpha$).

\section{Toomre Instability}
Self-gravity becomes important in disks outside of some radius, $r_{sg}$. When self-gravity becomes important, matter clumps together and will no longer accrete onto the compact object. The more massive the compact object, the closer $r_{sg}$ gets to the ISCO. When $r_{sg}=r_{ISCO}$ there will no longer be any accretion of luminous matter. Since accretion is the main mechanism by which supermassive black holes grow, this can be used to set an upper mass limit for supermassive black holes, as suggested in \citet{king}.\\

The radius at which self-gravity takes over can be determined by the Toomre parameter \citep{toomre} $Q=\frac{c_s\,\Omega_r}{\pi\,\Sigma}$, where $\Omega_r=\sqrt{\frac{-\frac{3 a^2}{r^2}+8 a \sqrt{M} r^{-3/2}-\frac{6 M}{r}+1}{r^3 \left(2 a \sqrt{M} r^{-3/2}-\frac{3 M}{r}+1\right)}}$ is the radial epicyclic frequency for circular, equatorial Kerr geodesics \citep{Gammie2004}. The condition for stability is that $Q>1$. Figure \ref{toomres} shows how $Q$ decreases with mass and approaches $1$ for masses $\sim 10^{11} M_{\odot}$. \\

We find that for the disk to remain Toomre stable ($Q>1$ everywhere outside the ISCO), we must have 
\be
M/M_{\odot}\leq f(a,\alpha) \sqrt{\dot{M}_{Edd}/\dot{M}},
\label{qineq}
\ee
where $f(0, 0.1)\approx 6.3\times 10^{21}$, $f(0.999, 0.1)\approx 1.1\times 10^{22}$, and $f(0, 0.01)\approx 6.3\times 10^{20}$. Restricting to sub-Eddington mass accretion rates $\dot{M}/\dot{M}_{Edd}\leq1$, together with the constraint from Toomre stability \eqref{qineq}, we get\\

\be
\begin{split}
\frac{dM}{dt}\leq \min\left[\dot{M}_{Edd}, \frac{\dot{{M}}_{Edd}(M_{\odot}) f(a,\alpha)}{M}\right]\\
\approx  \dot{{M}}_{Edd}(M_{\odot})\left(\frac{m^2}{f(a,\alpha)^2}+\frac{1}{m^2}\right)^{-1/2}.
\label{deq}
\end{split}
\ee

\begin{figure}[h]
\centering
\includegraphics[width=1.\textwidth]{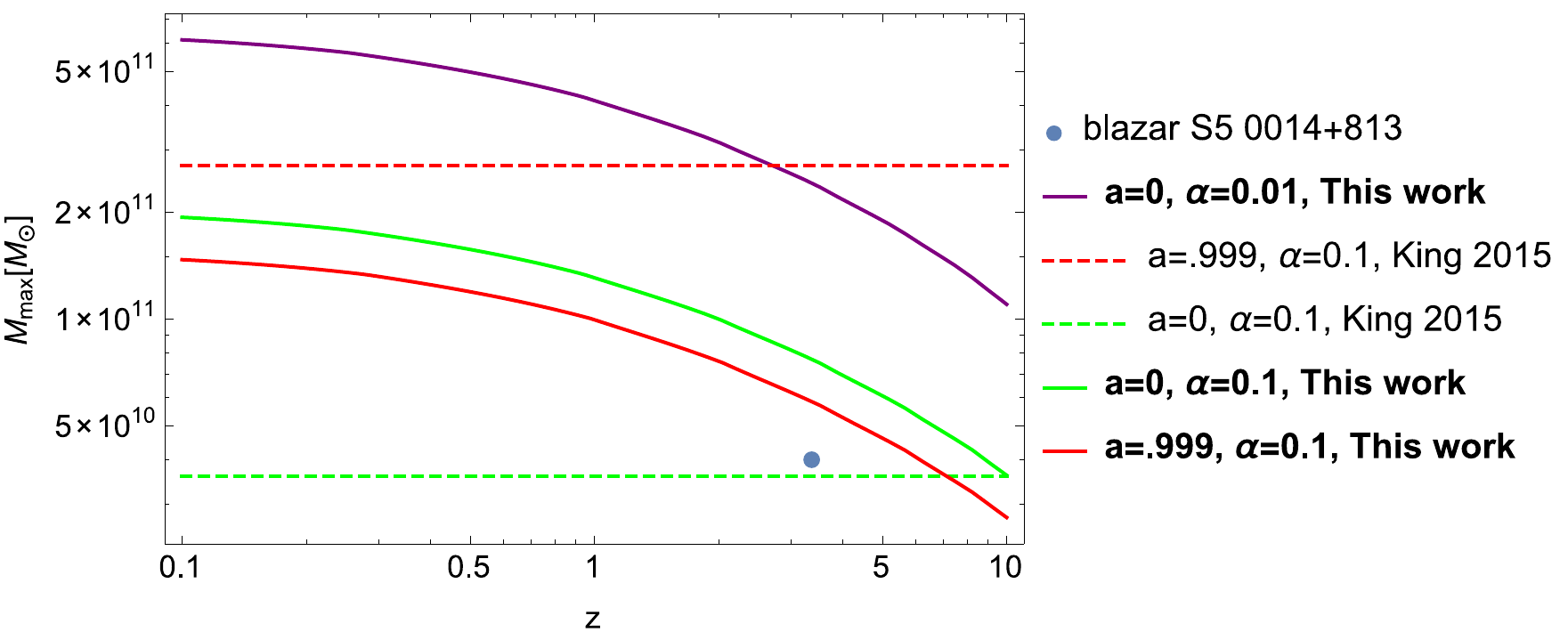}
\caption{Upper bounds for the mass of supermassive black holes at different redshifts $z$, due to the gravitational Toomre instability.}
\label{toomre}
\end{figure}

Integrating \eqref{deq} we have 
\be
t(M_f)\gtrsim \int_{M_i}^{M_f} dm \frac{1}{\dot{{M}}_{Edd}(M_{\odot})}\sqrt{\frac{m^2}{f(a,\alpha)^2}+\frac{1}{m^2}},
\ee

where masses and $t$ are in units of $M_{\odot}$. 
Inverting this relation to get $M_f(t)$, assuming that the first massive black holes were born at $z\sim 30$ and converting $t$ to $z$ to get $M(z)$, we arrive at the upper bounds shown in Figure \ref{toomre}, for nonspinning and nearly maximally spinning supermassive black holes. In the range of redshift shown in Figure \ref{toomre}, the mass bounds are not sensitive to the starting mass of the seed black hole ($M_i$). Also included in Figure \ref{toomre} are the upper bounds (for typical values of parameters) given in  \citet{king}, as well as a \emph{Swift} satellite observation of $S5 0014+813$ at $z=3.366$, where its mass was found to be $4\times 10^{10} M_{\odot}$ \citep{ghisellini1}. \\

\section{Conclusion and Future Prospects}
To summarize our main results,  we have introduced AiRE disks as a solution to thermal instability in thin disks. The key feature of AiRE disks is that the radiation pressure is in equipartition with the gas pressure in the inner region. We have presented some features of these flows such as their central temperature and Toomre parameter profiles. We have derived upper limits for the mass of supermassive black holes due to the gravitational Toomre instability in AiRE disks. We have also found a transition from saddle to nodal type of the sonic points in AiRE disks and used nodal point instability to place a lower limit on the mass accretion rate as a function of viscosity parameter $\alpha$ and black hole spin. We conjecture that this transition might be responsible for the observed lower limits on the Eddington ratio of the soft state in X-ray binaries. \\
 
While we introduced AiRE disks to provide a thermally stable description of thin accretion flows, they may also significantly refine our understanding of other disk properties. With new observations from advanced Laser Interferometer Gravitational-wave Observatory (aLIGO) and the Event Horizon Telescope (EHT), we are at the advent of a new era of black hole physics. Disk Theory may play a major role in explaining some of their future findings. \\

In upcoming work \citep{ykyna}, we study the spectrum of AiRE disks and its properties. Furthermore, the onset of Toomre instability in the inner regions of AiRE disks around active galactic nuclei can lead to formation and merger of  binary black hole systems, such as the ones recently detected by LIGO \citep{ligo}, and lead to characteristic detectable electromagnetic signatures  \citep[e.g.,][]{bartos}.\\

Another important future direction is the study of the AiRE disk regime in full MHD radiative transfer simulations, and whether enhanced cooling leading to pressure equipartition can indeed arise in a realistic setting.

\section*{Acknowledgments}
We would like to thank Shane Davis, Jeremy Goodman, Ramesh Narayan, Olek S{\c a}dowski, and Jim Stone for useful discussions. This research was supported in part by Perimeter Institute for Theoretical Physics. Research at Perimeter Institute is supported by the Government of Canada through the Department of Innovation, Science and Economic Development Canada and by the Province of Ontario through the Ministry of Research, Innovation and Science.

\bibliographystyle{mnras}

\label{lastpage}
\end{document}